\documentclass[twocolumn,aps,showpacs,pra,groupedaddress]{revtex4}
\usepackage{epsfig,amssymb,amsmath}
\usepackage[hypertex,linkcolor=red]{hyperref}
\def\comment#1{}\def\labell#1{\label{#1}}
\def\togliperprl#1{#1}
\begin{document}
\title{A quantum solution to the arrow-of-time dilemma}
\author{Lorenzo Maccone$^*$}\affiliation{QUIT, Dip.~A.~Volta, 27100
  Pavia, and Institute for Scientific Interchange, 10133 Torino,
  Italy.}

\begin{abstract}The arrow of time dilemma: the laws of physics are
  invariant for time inversion, whereas the familiar phenomena we see
  everyday are not (i.e.~entropy increases).  I show that, within a
  quantum mechanical framework, all phenomena which leave a trail of
  information behind (and hence can be studied by physics) are those
  where entropy necessarily increases or remains constant.  All
  phenomena where the entropy decreases must not leave any information
  of their having happened.  This situation is completely
  indistinguishable from their not having happened at all.  In the
  light of this observation, the second law of thermodynamics is
  reduced to a mere tautology: physics cannot study those processes
  where entropy has decreased, even if they were
  commonplace.\end{abstract}

\maketitle

Paradoxes have always been very fruitful in stimulating advances in
physics. One which still lacks a satisfactory explanation is the
Loschmidt paradox~\cite{los}. Namely, how can we obtain irreversible
phenomena from reversible time-symmetric physical laws~\cite{nota1}?
The irreversibility in Physics is summarized by the second law of
thermodynamics: entropy, which measures the degradation of the usable
energy in a system, never decreases in isolated systems. Many
approaches have been proposed to solve this conundrum, but most
ultimately resort to postulating low entropy initial states (see
e.g.~\cite{penrose,wald}), which is clearly an {\em ad hoc}
assumption~\cite{antih}. Others suggest that the thermodynamic arrow
of time is in some way connected to the cosmological
one~\cite{davies}, that physical laws must be modified to embed
irreversibility~\cite{gianpa}, that irreversibility arises from
decoherence~\cite{zeh}, or from some time-symmetric mechanism embedded
in quantum mechanics~\cite{schulm}, {\em etc.}  Recent reviews on this
problem are given in Ref.~\cite{reviews}.

Here I propose a different approach, based on existing laws of physics
(quantum mechanics). I show that entropy in a system can both increase
and decrease (as time reversal dictates), but that all
entropy-decreasing transformations cannot leave {\em any} trace of
their having happened. Since no information on them exists, this is
indistinguishable from the situation in which such transformations do
not happen at all: ``The past exists only insofar as it is recorded in
the present''~\cite{wheel}. Then the second law is forcefully valid:
the only physical evolutions we see in our past, and which can then be
studied, are those where entropy has not decreased.

I start by briefly relating the thermodynamic entropy with the von
Neumann entropy, and introducing the second law. I then present two
thought experiments, where entropy is deleted together with all
records of the entropy increasing processes: even though at some time
the entropy of the system had definitely increased, afterward it is
decreased again, but none of the observers can be aware of it. I
conclude with a general derivation through the analysis of the entropy
transfers that take place in physical transformations.

\paragraph*{Entropy and the second law.}
Thermodynamic entropy is a quantity that measures how the usable
energy in a physical process is degraded into heat. It can be
introduced in many ways from different axiomatizations of
thermodynamics. The von Neumann entropy of a quantum system in the
state $\rho$ is defined as $S(\rho)\equiv-$Tr$[\rho\log_2\rho]$. When
applicable, these two entropies coincide (except for an
inconsequential multiplicative factor). This derives from an argument
introduced by Einstein~\cite{einstein} and extended by
Peres~\cite{peres} (e.g.~both the canonical and the microcanonical
ensemble can be derived from quantum mechanical
considerations~\cite{popescu,can}).  For our purposes, however, it is
sufficient to observe that thermodynamic and von Neumann entropies can
be inter-converted, employing Maxwell-demons~\cite{demon,seth1} or
Szilard-engines~\cite{szilard,scully}: useful work can be extracted
from a single thermal reservoir by increasing the von Neumann entropy
of a memory space.

There are many different formulations of the second law, but we can
summarize them by stating that, in any process in which an isolated
system goes from one state to another, its thermodynamic entropy does
not decrease~\cite{thermo}. There is a hidden assumption in this
statement.  Whenever an isolated system is obtained by joining two
previously isolated systems, then the second law is valid only if the
two systems are initially uncorrelated, i.e.~if their initial joint
entropy is the sum of their individual entropies. It is generally
impossible to exclude that two systems might be correlated in some
unknown way and there is no operative method to determine whether a
system is uncorrelated from all others (e.g.~given a box containing
some gas it is impossible to exclude that the gas particles might be
correlated with other systems). Thus, in thermodynamics all systems
are considered uncorrelated, unless it is known otherwise.  Without
this assumption, it would be impossible to assign an entropy to any
system unless the state of the whole universe is known: a normal
observer is limited in the information she can acquire and on the
control she can apply. This implies that thermodynamic entropy is a
subjective quantity\togliperprl{~\cite{sethent}}, even though {\em for
  all practical situations} this is completely irrelevant: the
eventual correlations in all macroscopic systems are practically
impossible to control and exploit. Even though they are ignored by the
normal observer, correlations between herself and other systems do
exist.  Until they are eliminated, the other systems cannot decrease
their entropy. A physical process may either reduce or increase these
correlations.  When they are reduced, this may seem to entail a
diminishing of the entropy, but the observer will not be aware of it
as her memories are correlations and will have been erased by
necessity (each bit of memory is one bit of correlation and, until her
memory has been erased, the correlations are not eliminated). Instead,
when the physical process increases these correlations, she will see
it as an increase in entropy. The observer will then only be aware of
entropy non-decreasing processes. \togliperprl{[Not even a
  super-observer that can keep track of all the correlations would
  ever see any entropy decrease. In fact, since he can discover and
  take advantage of all correlations between microscopic degrees of
  freedom, all processes are always zero-entropy processes from his
  super-observer point of view.]}

The above analysis is limited to systems that are somehow correlated
with the observer. One might then expect that she could witness
entropy decreasing processes in systems that are completely factorized
from her. That is indeed the case: statistical microscopic
fluctuations can occasionally decrease the entropy of a system (the
second law has only a statistical valence). However, an observer is
macroscopic by definition, and all remotely interacting macroscopic
systems become correlated very rapidly (e.g.~Borel famously calculated
that moving a gram of material on the star Sirius by one meter can
influence the trajectories of the particles in a gas on earth on a
time-scale of $\mu$s~\cite{borel}).  This is the same mechanism at
the basis of quantum decoherence~\cite{zeh}, and it entails that in
practice the above analysis applies to all situations: no entropy
decrease in macroscopic systems is ever observed.

In what follows I will make these ideas rigorous.

\togliperprl{Since the two above definitions of entropy are
  equivalent, the von Neumann entropy also obeys the second law. In
  fact, isolated systems evolve with unitary evolutions, which leave
  the von Neumann entropy invariant.  There may be an increase if the
  evolution is not exactly known or if it creates unknown correlations
  among subsystems. In the first case, the coarse-grained evolution is
  of the form $\rho^\prime=\sum_np_nU_n\rho {U_n}^\dag$, where $p_n$
  is a probability and $U_n$ are unitary operators. Then the final
  entropy $S(\rho^\prime)$ may be larger than the initial entropy
  $S(\rho)$:
\begin{eqnarray}
  S(\rho^\prime)=S(\sum_np_nU_n\rho {U_n}^\dag)\geqslant
  \sum_np_nS(U_n\rho {U_n}^\dag)=S(\rho)\labell{unk}
\end{eqnarray} 
(the inequality follows from the concavity of the entropy). In the
second case, the entropy of subsystems can increase, as some unknown
correlations between them may build up:
\begin{eqnarray}
  S(\rho^\prime_1)+S(\rho^\prime_2)\geqslant
  S(U(\rho_1\otimes\rho_2)U^\dag)=
  S(\rho_1)+S(\rho_2)\;,
\end{eqnarray} 
where $\rho^\prime_i$ and $\rho_i$ are the final and initial states of
the subsystems, $U$ is the evolution coupling them, and the last
equality holds if they are initially uncorrelated (the
  inequality follows from the subadditivity of the entropy).}

\paragraph*{Thought experiments.}
The quantum information theory mantra ``Information is
physical''~\cite{landauer} implies that any record~\cite{nota2} of an
occurred event can be decorrelated from such event by an appropriate
physical interaction.  If {\em all} the records of an event are
decorrelated from it, then by definition there is no way to know
whether this event has ever happened. This situation is
indistinguishable from its not having happened. If this event has
increased the entropy, the subsequent erasing of all records can
($\!${\em will}) produce an entropy decrease without violation of any
physical law. We now analyze two such situations, an imperfect
transmission of energy and a quantum measurement.

Alice's lab is perfectly isolated, so that to an outside observer
(Bob), its quantum evolution is unitary. Analyze the situation in
which Bob sends Alice some energy in the form of light, a multimode
electromagnetic field in a zero-entropy pure state.  We suppose that,
to secure the energy Bob is sending her, she uses many detectors which
are not matched to his modes. Given a system in almost any possible
pure state, all its subsystems which are small enough are
approximately in the canonical state~\cite{popescu}.  This implies
that, if each of Alice's detectors is sensitive to only a small part
of Bob's modes, the detectors mostly see thermal radiation, and she
feels them warming up.  She will then be justified in assigning a
nonzero thermodynamic entropy to her detectors, as she sees them
basically as thermal-equilibrium systems.  One might object that she
is mistaken, since the states of the detectors are not uncorrelated.
However, since she ignores the correlations, she cannot use such
correlations to extract energy from the detectors. Alice concludes
that most of the energy Bob sent her has been wasted as heat, raising
the thermodynamic entropy of her lab.  Suppose now that Bob has
complete control of all the degrees of freedom in her lab. He knows
and can exploit the correlations to recover all the energy he had
initially given Alice.  Of course, although possible in principle, he
needs a dauntingly complex transformation, which requires him to be
able to control a huge number of her lab's degrees of freedom
(including the brain cells where her memories are, and the notepads
where she wrote the temperatures!). To extract the energy, since it
was initially locked in a pure state of the field, he must return it
to a system in a zero-entropy pure state, i.e.~factorized from all the
other degrees of freedom of Alice's lab. Then he must erase {\em all}
the correlations between them: at the end of Bob's recovery, Alice
cannot remember feeling her detectors warm up, they are cool again,
her notepads contain no temperature information, and all the energy
initially in the electromagnetic field is again available, even though
(from Alice's point of view) most of it was definitely locked into
thermal energy at one time.

\togliperprl{
\begin{figure}[t]
\begin{center}
\epsfxsize=.8\hsize\leavevmode\epsffile{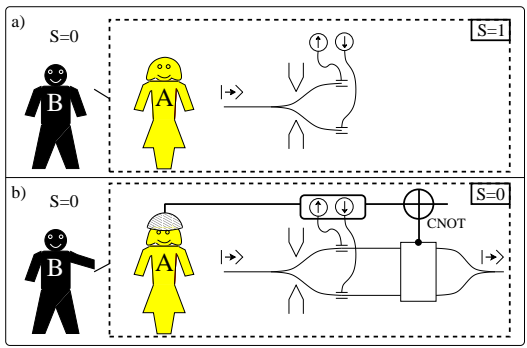} 
\end{center}
\caption{a) Alice in her isolated lab performs a Stern-Gerlach
  measurement on a spin 1/2 particle initially oriented parallel to
  the $x$ axis, i.e.~in a state $|\rightarrow\rangle$. Since the
  apparatus is oriented along the $z$ axis, this measurement creates
  one bit of entropy for Alice (not for Bob, who is isolated from her
  lab). b) Bob flips a switch that ``cancels'' Alice measurement by
  decorrelating from the spin all those degrees of freedom of her lab
  that have recorded the measurement outcome. Now the spin is returned
  to its initial state $|\rightarrow\rangle$ and Alice cannot have any
  memory of what her measurement result was. Her entropy has
  decreased, but she cannot remember it ever having increased.}
\labell{f:lab}
\end{figure}
}

The second though-experiment~\cite{everett} is a prototypical quantum
measurement.  Bob prepares a spin-1/2 particle oriented along the $x$
axis, e.g.~in a spin $|\rightarrow\rangle$ state and hands it to
Alice. She sends it through a Stern-Gerlach apparatus oriented along
the $z$ axis~\cite{peres}. The measurement consists in coupling the
quantum system with some macroscopic degrees of freedom (a reservoir),
not all of which are under the control of the
experimenter~\cite{peres2}, whence the irreversibility. Notice that,
$|\rightarrow\rangle=(|\uparrow\:\rangle+|\downarrow\:\rangle)/\sqrt{2}$,
where $|\uparrow\:\rangle$ and $|\downarrow\:\rangle$ are the
eigenstates of a $z$ measurement operator. Hence, this apparatus will
increase the entropy of the spin system by one bit~\cite{seth1}:
Before the readout, the spin state will be in the maximally mixed
state $(|\uparrow\:\rangle\langle\:\uparrow|+
|\downarrow\:\rangle\langle\:\downarrow|)/2$. After Alice has looked
at the result, she has transferred this one bit of entropy, created by
the measurement, to her memory. From the point of view of Bob, outside
her isolated lab, Alice's measurement is simply a (quantum)
correlation of her measurement apparatus to the spin. [A thorough
analysis of the microscopic details and of the thermodynamics of this
type of measurement is given in Ref.~\cite{seth1}.] The initial state
of the spin
$|\rightarrow\rangle=(|\uparrow\:\rangle+|\downarrow\:\rangle)/\sqrt{2}$
evolves into the correlated (entangled) state
\begin{eqnarray}
  (|\uparrow\:\rangle|\mbox{Alice sees
    ``up''}\rangle+|\downarrow\:\rangle|\mbox{Alice sees
    ``down''}\rangle)/\sqrt{2}
\labell{la}\;,
\end{eqnarray}
where the first ket in the two terms refers to the spin state, whereas
the second ket refers to the rest of Alice's lab. Thus, from the point
of view of Bob, Alice's measurement is an evolution similar to a
controlled-NOT unitary transformation of the type
$U_{cnot}(|0\rangle+|1\rangle)|0\rangle=|0\rangle|0\rangle+|1\rangle|1\rangle$.
Such a transformation can be easily inverted, as it is its own
inverse.  Analogously, Bob can flip a switch and invert Alice's
measurement.  At the end of his operation, all records of the
measurement result (Alice's notepad, her brain cells, the apparatus
gauges) will have been decorrelated from the spin state.  She will
remember having performed the measurement, but she will be ({\em must}
be) unable to recall what the measurement result was. In addition, the
spin has become uncorrelated from the measurement apparatus, so it is
returned to a pure state.  I emphasize that Bob's transformation is
not necessarily a reversion of the dynamics of Alice's lab.
\togliperprl{[Notice that a ``quantum eraser''~\cite{erasure} only permits
  to decide {\em a posteriori} which of two complementary measurements
  to perform using previously collected data: the measurement process
  is not actually erased, and the entropy does not decrease.]}

In both the above experiments, from Alice's point of view, entropy
definitely has been created after she has interacted with Bob's light
or his spin.  However, this entropy is subsequently coherently erased
by Bob. At the end of the process, looking back at the evolution in
her lab, she cannot see any violation of the second law: she has no
(cannot have any) record of the fact that entropy at one point had
increased.

\paragraph*{Entropic considerations.}
The above thought experiments exemplify a general situation: entropy
can decrease, but its decrease is accompanied by an erasure of any
memory that the entropy-decreasing transformation has occurred.  In
fact, any interaction between an observer $A$ and a system $C$ which
decreases their entropy by a certain quantity, must also reduce their
quantum mutual information by the same amount (unless, of course, the
entropy is dumped into a reservoir $R$). The quantum mutual
information $S(A:C)\equiv S(\rho_A)+S(\rho_C)-S(\rho_{AC})$ measures
the amount of shared quantum correlations between the two systems $A$
and $C$ ($\rho_{AC}$ being the state of the system $AC$, and $\rho_A$
and $\rho_C$ its partial traces, i.e.~the states of $A$ and $C$).

Taking the cue from~\cite{seth}, I now prove the above assertion,
namely I show that 
\begin{eqnarray}
\Delta S(A)+\Delta S(C)-\Delta
S(R)-\Delta S(A:C)=0\labell{th}\;,
\end{eqnarray}
where $\Delta S(X)\equiv S_t(\rho_X)-S_0(\rho_X)$ is the entropy
difference between the final state at time $t$ and the initial state
of the system $X$, and where $\Delta S(A:C)=S_t(A:C)-S_0(A:C)$ is the
quantum mutual information difference. Choose the reservoir $R$ so
that the joint state of the systems $ACR$ is pure and so that the
evolution maintains the purity ($R$ is a purification space). Then the
initial and final entropies are $S_0(AC)=S_0(R)$ and $S_t(AC)=S_t(R)$,
respectively. Thus we find $S_0(AB)=S_t(AB)-\Delta S(R)$ which, when
substituted into the left-hand-side term of (\ref{th}), shows that
this term is null. [This proof is valid also if the evolution is not
perfectly known, i.e. if it is given by a random unitary
map\togliperprl{, see Eq.~(\ref{unk})}.]

Now, to prove that the above reduction of entropy entails a memory
erasure, I show that this erasure must follow from the elimination of
quantum mutual information.  A memory of an event is a physical system
$A$ which has nonzero classical mutual information on a system $C$
that bears the consequences of that event.  Then, the erasure of the
memory follows from an elimination of the quantum mutual information
$S(A:C)$ if this last quantity is an upper bound to the classical
mutual information $I(A:C)$. Thus, we must show that for any POVM
measurement $\{\Pi^{(a)}_i\otimes\Pi^{(c)}_j\}$ extracting information
separately from the two systems ($\Pi^{(a)}_i$ acting on $A$ and
$\Pi^{(c)}_i$ on $C$), \comment{Nel paper \cite{ozawayuen}, la
  dimostrazione dell'equazione (12) e' praticamente la mia
  dimostrazione qui sotto, anche se l'equazione (12) non puo'
  immediatamente essere collegata con la mutua informazione per come
  e' introdotto $\rho_\theta$ li'}
\begin{eqnarray}
S(A:C)\geqslant I(A:C)
\labell{th2}\;,
\end{eqnarray}
where $I(A:C)$ is the mutual information of the POVM's measurement
results. A simple proof of this statement exists
(e.g.~see~\cite{ozawayuen,barbara}): use the equality
$S(A:C)=S(\rho_{AC}||\rho_A\otimes\rho_C)$, where
$S(\rho\|\sigma)\equiv$Tr$[\rho\log_2\rho-\rho\log_2\sigma]$ is the
quantum relative entropy. This quantity is monotone for application of
CP-maps~\cite{ozawayuen}, i.e.~$S(\rho\|\sigma)\geqslant S({\cal
  N}[\rho]\|{\cal N}[\sigma])$ for any transformation $\cal N$ that
can be written as ${\cal N}[\rho]=\sum_kA_k\rho {A_k}^\dag$, with the
Kraus operators $A_k$ satisfying $\sum_k{A_k}^\dag A_k=\openone$.
Consider the ``measure and reprepare'' channel, i.e.  the
transformation ${\cal
  N}[\rho]=\sum_n\mbox{Tr}[\Pi_n\rho]\:|n\rangle\langle n|$ where
$\{|n\rangle\}$ is a basis, and $\Pi_n$ is a POVM element (i.e.~a
positive operator such that $\sum_n\Pi_n=\openone$). It is a CP-map,
since it has a Kraus form
\begin{eqnarray}
A_{nm}=|n\rangle\langle v_m^{(n)}|\sqrt{}p_m^{(n)}\mbox{, with }
\Pi_n=\sum_mp_m^{(n)}|v_m^{(n)}\rangle\langle v_m^{(n)}|
\labell{kraus}\;.
\end{eqnarray}
Using the monotonicity of the relative entropy under the action of the
map $\cal N$, we find
\begin{eqnarray}
&S(A:C)=S(\rho_{AC}||\rho_A\otimes\rho_C)
\geqslant S({\cal N}[\rho_{AC}]\:\|\:{\cal N}[\rho_A\otimes\rho_C])&
\nonumber\\\nonumber&=
\sum_{ij}p_{ij}\log_2p_{ij}-\sum_{ij}p_{ij}\log_2(q_ir_j)=I(A:C)
\;,&
\end{eqnarray}
where $p_{ij}\!\equiv$Tr$[\Pi^{(a)}_i\otimes\Pi^{(c)}_j\rho_{AC}]$,
$q_{i}\!\equiv$Tr$[\Pi^{(a)}_i\rho_{A}]$, and
$r_{j}\!\equiv$Tr$[\Pi^{(c)}_j\rho_{C}]$.  

The interpretation of Eq.~(\ref{th}) is that, if we want to decrease
the entropy of the system $C$ (somehow correlated with the observer
$A$) without increasing the entropy of a reservoir $R$, we need to
reduce the quantum mutual information between $C$ and the
observer $A$\togliperprl{ (e.g.~in the Stern-Gerlach thought experiment,
  the system $A$ is Alice's lab and $C$ is the spin-1/2 particle:
  their final entropies are reduced by one bit at the expense of
  erasing two bits of quantum mutual information $S_0(A:C)$)}. The
fact that mutual information can be used to decrease entropy was
already pointed out by Lloyd~\cite{seth} and Zurek~\cite{zu}.

The implications of the above analysis can be seen explicitly by
employing Eq.~(\ref{th}) twice, by considering an intermediate time
when $S(C)$ is higher than at the initial and final times. The entropy
$S(C)$ of the system is high at the intermediate time after an
entropy-increasing transformation, and then (if no entropy-absorbing
reservoir $R$ is used) it can be reduced by a successive
entropy-decreasing transformation at the cost of reducing the mutual
information between the observer and $C$.  Even though the entropy
$S(C)$ (as measured from the point of view of the observer $A$) does
decrease, the observer is not aware of it, as the entropy-decreasing
transformation must factorize her from the system $C$ containing
information on the prior entropy-increasing event: her memories of
such event must be part of the destroyed correlations.
\togliperprl{The deep reason for this is that, from her own point of
  view, the Born rule kicks in when the observer becomes entangled
  with another system.  (An external super-observer may, instead, just
  see her becoming entangled with the other system, but then he cannot
  know the measurement result.) The Born rule is the only place where
  quantum mechanics allows irreversibility, but the correlations that
  stem from such rule {\em can} be undone, at least if one treats both
  the observer and the apparatus quantum mechanically. This means that
  the measurement can be undone, at the price that all the observer's
  memories must be erased.}

What we have seen up to now is that any decrease in entropy {\em of a
  system that is correlated with an observer} entails a memory erasure
of said observer, in the absence of reservoirs (or is a zero-entropy
process for a super-observer that keeps track of all the
correlations). That might seem to imply that an observer should be
able to see entropy-decreasing processes when considering systems that
are uncorrelated from her. In fact, at microscopic level, statistical
fluctuations do decrease occasionally the entropy. However, the
correlations between any two macroscopic systems build up
continuously, and at amazing rates~\cite{borel}: this is how
decoherence arises~\cite{zeh}.  Then no observer is really factorized
with respect to any macroscopic system she observes. This implies that
entropy decreases of a macroscopic system becomes unobservable (unless
extreme care is taken to shield the system under analysis). Only
microscopic systems can be considered factorized from an observer for
a period of time long enough to see entropy decrease from 
fluctuations.

\paragraph*{Conclusions.}
In this paper I gave a quantum solution to the Loschmidt paradox,
showing that all physical transformations where entropy is decreased
cannot relinquish any memory of their having happened from the point
of view of any observer: both normal observers that interact with the
studied systems and external super-observers that keep track of all
the correlations. Thus they are irrelevant to physics.  Quantum
mechanics is necessary to this argument. In the above derivation, we
have used the property that the entropy of a joint system can be
smaller than that of each of its subsystems.  This is true of von
Neumann entropy, but not true if entropy is calculated using classical
probability theory: then the entropy of a joint system is always
larger than that of its subsystem with largest entropy.
\togliperprl{By how much must any system be extended until we can take
  advantage of this quantum reduction of the global entropy?  It is
  clear from Borel's famous arguments~\cite{borel} that the time scale
  in which a macroscopic system can be really considered as isolated
  is very small. As such, the arguments presented in this paper are of
  theoretical interest only, and have little or no practical
  consequence for any normal macroscopic system: the effects presented
  here become relevant only at a scale that approaches the whole
  universe very rapidly.}

\togliperprl{In closing, I indulge in a couple of more philosophical
  considerations.}  In a quantum cosmological setting, the above
approach easily fits in the hypothesis that the quantum state of the
whole universe is a pure (i.e.~zero-entropy) state evolving unitarily
(e.g.~see~\cite{max,albert2,popescu}).  One of the most puzzling aspects of
our universe is the fact that its initial state had entropy so much
lower than we see today, making the initial state highly
unlikely~\cite{wald}. Joining the above hypothesis of a zero-entropy
pure state of the universe with the second law considerations analyzed
in this paper, it is clear that such puzzle can be resolved.  The
universe may be in a zero entropy state, even though it appears (to
us, internal observers) to possess a higher entropy\togliperprl{: our
  situation is similar to the one of Alice, who, just after the
  measurement sees her lab in a nonzero entropy state, whereas to the
  super-observer Bob her lab maintains a zero-entropy state all
  along}.  However, it is clear that this approach does not require to
deal with the quantum state of the whole universe, but it applies also
to arbitrary physical systems.

\togliperprl{In a quantum cosmological framework, Boltzmann's initial
  condition translates in the equivalent question of why the initial
  state of the universe is such that its subsystems are mostly
  unentangled. Such a state is highly improbable, as all states in a
  sufficiently large Hilbert space are almost completely
  entangled~\cite{qqq}. The most compelling answer to this question
  derives from Davies' argument that, as space expands because of
  cosmological expansion and inflation, new degrees of freedom are
  created, giving the potential for accommodating new
  entropy~\cite{davies1}: one may think that, as these new degrees of
  freedom are created, they are initialized in a factorized (possibly
  pure) state. However, this assumption is unwarranted, and moreover,
  our current understanding of quantum mechanics does not allow the
  description of a situation where the number of degrees of freedom of
  a system (and hence its Hilbert space) changes
  dynamically~\cite{unruh}. The alternative solution presented in this
  paper sidesteps the problem: whatever the state of the universe, an
  internal observer would still only see the processes where entropy
  increases. }

\togliperprl{In addition, I recall that there is a substantial problem
  in rigorously defining past and future without resorting to the
  second law (which would then be reduced to a mere definition). In
  fact, the laws of physics are time-reversal invariant. Hence, there
  is no preferred direction of time according to which we may
  establish a {\em substantial} difference between the two temporal
  directions past-to-future and future-to-past~\cite{castagnino}.
  Anthropocentrically, we could define the past as that of which we
  have memories of, and the future as that of which we do not have any
  memories. Of course, such definition cannot be made rigorous, since
  it resorts to a observers and their memories. However, even using
  this ambiguous, intuitive definition of past, it is clear that any
  event, which cannot have any correlation with us, does not pertain
  to our past just as if it had never happened.}

I thank S. Lloyd, G.M. D'Ariano, V. Giovannetti, G.P.  Beretta, S.
Wolf, and A.  Winter for useful discussions and feedback.

\vskip 1\baselineskip
{\footnotesize * Current address: Massachusetts Institute of
Technology, RLE, 50 Vassar Street Bldg. 36-472B
Cambridge, MA 02139 USA.}


\begin{references}
\bibitem{nota2} By record I consider any physical system which is
  correlated with the event, such as its description on the pages of a
  book, or the neural pattern in a witness's brain.
\end{references}
\end{document}